\documentclass[conference]{IEEEtran}
\IEEEoverridecommandlockouts
\usepackage{hyperref}
\hypersetup{colorlinks,allcolors=black}
\usepackage{cite}
\usepackage{amsmath,amssymb,amsfonts}
\usepackage{algorithmic}
\usepackage{graphicx}
\usepackage{textcomp}
\usepackage{xcolor}
\usepackage{float}
\usepackage{diagbox}
\usepackage{soul}

\makeatletter
\newcommand{\linebreakand}{%
  \end{@IEEEauthorhalign}
  \hfill\mbox{}\par
  \mbox{}\hfill\begin{@IEEEauthorhalign}
}
\makeatother
  
\def\BibTeX{{\rm B\kern-.05em{\sc i\kern-.025em b}\kern-.08em
    T\kern-.1667em\lower.7ex\hbox{E}\kern-.125emX}}
\begin{document}

\title{The Effect of Pore Structure in Flapping Wings on Flight Performance}

\author{\IEEEauthorblockN{1\textsuperscript{st} Abdurrahim Yilmaz}
\IEEEauthorblockA{\textit{Yildiz Technical University} \\
\textit{Mechatronics Engineering}\\
Istanbul, Turkey \\
a.rahim.yilmaz@gmail.com}
\and
\IEEEauthorblockN{2\textsuperscript{nd} Asli Tekeci}
\IEEEauthorblockA{\textit{Yildiz Tehnical University} \\
\textit{Mechatronics Engineering}\\
Istanbul, Turkey \\
asli.tekeci.37@gmail.com}
\and
\IEEEauthorblockN{3\textsuperscript{rd} Meryem Ece Ozyetkin}
\IEEEauthorblockA{\textit{Yildiz Tehnical University} \\
\textit{Mechatronics Engineering}\\
Istanbul, Turkey \\
eceozyetkin@gmail.com}
\linebreakand
\IEEEauthorblockN{4\textsuperscript{th} Ali Anil Demircali}
\IEEEauthorblockA{\textit{Imperial College London} \\
\textit{The Hamlyn Centre}\\
London, The United Kingdom \\
a.demircali@imperial.ac.uk}
\and
\IEEEauthorblockN{5\textsuperscript{th} Kagan Unsal}
\IEEEauthorblockA{\textit{SEV American College} \\
Istanbul, Turkey \\
kunsal22@my.sevkoleji.k12.tr}
\and
\IEEEauthorblockN{6\textsuperscript{th} Huseyin Uvet}
\IEEEauthorblockA{\textit{Yildiz Tehnical University} \\
\textit{Mechatronics Engineering}\\
Istanbul, Turkey \\
huvet@yildiz.edu.tr}
}

\maketitle

\begin{abstract}

This study investigate the effects of porosity on an flying creatures such as dragonfly, moths, hummingbird etc. wing and shows that pores can affect wing performance. These studies were preformed by 3D porous flapping wing flow analyses on Comsol Multiphysics. In this study, we analyzed different numbers of porous wing at different angles of inclination in order to see the effect of pores on lift and drag forces. To compare the results 9 different analyses were performed. In these analyses, air flow velocity was taken as 5 m/s, angle of attack as 5 degrees, frequency as 25 Hz, and flapping angle as 30 degree. By keeping these values constant, the number of pores was changed to 36, 48, and 60 and the pore angles of inclination to 60, 70, and 80 degrees. Analyses were carried out by giving laminar flow to this wing designed in the Comsol Multiphysics program. The importance of pores was investigated by comparing the results of these analyses.

\end{abstract} 

\begin{IEEEkeywords}
Micro Air Vehicle (MWA), Flapping Wing, Fluid Multibody Interaction, Lift and Drag Force, COMSOL Multiphysics, Porous Wing
\end{IEEEkeywords}

\section{Introduction}

Unmanned Aerial Vehicle (UAV) is a multi-purpose aircraft that operates remotely or autonomously, they are multi-purpose vehicles which range in size from the size of a passenger plane to the size of an insect. Micro air vehicle (MAV) that are one of UAV types has been the argument of many researches and developed in recent years. As a result of biomimetic (imitating nature) researches, small birds and insects are good objects to be imitated to design MAV, because they perform stable flight at low Reynolds number. Although, there is no precise definition for micro aircraft, according to the definition of DARPA, The Defense Advanced Research Projects Agency, aircraft that do not exceed 15 centimeters in length are classified as MAV \cite{mcmichael1997micro}. Flapping wing micro air vehicles (FWMAVs) have several advantages over other UAVs due to their maneuverability, durability and high power efficiency at low flight speeds. FWMAVs have features such as maneuvering, avoiding obstacles, gliding, soaring, flying at low speeds and narrow environments, short-distance takeoffs and landings, and hovering in the air. Because they have various flight characteristics, FWMAVs can adapt to changing aerodynamic conditions. They can also vary their angle of attack (AOA), wing area, and flapping frequency \cite{Yang2018THETF}. These features have led us to investigate the wing flapping kinematics of flying creatures and to analyze the effect of pores on their wings in order to increase the flight performance of FWMAVs.

There are some problems and deficiencies that still need to be solved in the current MAVs. One of these problems is wing performance. Researchers started biological studies to solve this problem inspired by nature. In nature, flight forces are usually obtained by flapping wings. In the last decade, based on this knowledge, many researchers have been inspired by insects that can fly in nature to improve MAVs. Some of these studies are Manduca Sexta FWMAV \cite{moses2017artificial}, Micro-mechanical Flying Insects \cite{deng2006flapping}, Dragonfly \cite{azuma1985flight}, Butterfly \cite{sunada1993performance}, RoboBee \cite{wood2008first}, RoboFly \cite{singh2008insect}, KUBeetle \cite{phan2017design}, FlowerFly \cite{nguyen2015performance}, DelFly \cite{bruggeman2010improving}, H2bird \cite{rose2015modeling}, Microbat \cite{pornsin2000microbat}, Golden Snitch \cite{yang2012micro}, Bat Bot \cite{ramezani2016bat}, Hummingbird \cite{roshanbin2017colibri}, and SmartBird \cite{corporate2012smartbird}. As mentioned in such studies, wing performances such as wing beat frequency, wing surface, number of pores on wing surface, aspect ratio, lift coefficient and wing loading are effective in flying creatures that can fly in nature. These can cause differences in wing beating kinematics \cite{willmott1997mechanics}. The researchers studied various bird and insect wing types. As an illustration, P. Wu et al. studied the flapping wing kinematics of different flying creatures \cite{wu2011structural}. As a result of this study, it was observed that the frequency range and wing flexibility were effective in thrust force. In addition, rigid and flexible wings have been tested and it has been concluded that the performance of rigid wings is worse than flexible wings \cite{pelletier2000low} \cite{mazaheri2010experimental}.

Computational fluid dynamics (CFD) analysis have been used in numerous studies in order to search flapping wing using two-dimensional (2D) wing profiles \cite{kumar2016analysis}. Also there are few studies on the three-dimensional (3D) CFD wing analysis of FWMAVs. Due to the time-varying and flexible nature of the fluttering wing, 2D analysis is not sufficient, so 3D analysis has been performed. It was decided that COMSOL Multi-physics, which has Fluid Structure Interaction (FSI) feature. Because it is more powerful than others in the way of calculation among the programs where this analysis can be performed. In addition, the effect of pores on the wings of flying creatures such as moth, hummingbird inspired in this study on lift-drag forces produced during flight was wondered. For these reasons, in our study that on flapping wings, we performed a numerical simulation of the 3D wing beat flow using COMSOL Multi-physics to approximate the lift-drag forces that occur in flexible wings with varying pore numbers. In other words, from these analyses results, we observed how lift and drag forces are affected by changing wing conditions such as the number of pores in the wing.

\section{Methodology}

Two of the important features when designing wings of MAVs are lift and drag forces. In this study, wing characteristics of various flying creatures were investigated in order to design the wings of MAVs. Some of wing properties of flapping wing are frequency, wing dimensions etc. shown in Table \ref{tab:p1}. The effects of the number of pores and inclination angle of pores in the wings of flying creature on lift and drag forces were analyzed.

\begin{table}[ht]
\renewcommand{\arraystretch}{1.2}
\centering
\caption{WING PROPERTIES}
 \begin{tabular}{|| c c ||}
 \hline \hline
 Species & Values  \\ 
 \hline
Frequency (Hz) & 25 \\
\hline
Wing Dimensions (cm) & 10x10x1\\
\hline
Wing Aspect Ratio & 1 \\
\hline
Wing Material & SU8-photoresist \\
\hline\hline 
\end{tabular}
\label{tab:p1}
\end{table}

\subsection {Model Definition}
After investigating the properties of flying creatures, the flapping wing was designed to analysis by inspiring from these features. The model geometry of wing consists of a mechanism with a wing and a cylinder which used for wing beat. This flapping wing model is 100 mm wide, 100 mm deep and 10 mm high. The cylinder is assembled at the tip of the wing and it connected to the wing through hinge joints, which let in-plate rotation of the cylinder. Also, wing has various number of pores in the wing, such as 36, 48 and 60. Analyses with these pore numbers were made when the pores were at different angles of inclination. The angle of inclination of these pores varies as 60, 70 and 80 degrees. Thus, 9 analyses were made. The mechanism is in a wind tunnel of 400 mm width, 500 mm depth, and 500 mm height. In this wind tunnel, the mechanism is positioned with an angle of attack of 5 degrees. In this analysis model assemblies materials are combination of flexible and rigid components in this reason The Multibody Dynamics interface was used in the Comsol Multiphysics. In order to model the flapping wing mechanism, the joints, attachments etc. properties of this interface were used.

\begin{figure}[ht]
    \centering
    \includegraphics[width=0.5\textwidth]{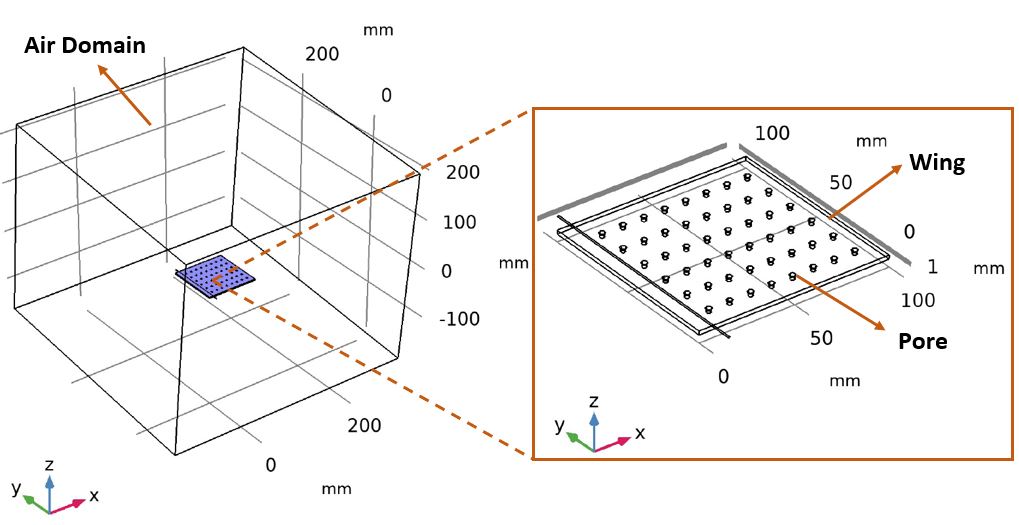}
    \caption{Example geometry in a wind tunnel}
    \label{fig:p1}
\end{figure}

The designed mechanism was analysed with the Fluid-Multibody Interaction solution module that in the Comsol Multiphysics program, as shown in Figure \ref{fig:p1}.

\subsection {Fluid-Multibody Interaction}
Fluid-Multibody Interaction solution module is a multiphysics problem that used to model the affect to each other of an assembled solid mechanism and a fluid. This module falls into the fluid-structure interaction (FSI) class.

These analyses show the dynamics of a flapping wing mechanism placed into a wind tunnel. The fluid, which is air, conforms to the laminar flow pattern. This type of flow moves uninterrupted and parallel layers within the flow area where it is located. This wing mechanism is modeled using the Multibody Dynamics interface. A Fluid Structure Interaction and Pair Multiphysics Coupling was used to model the effects of solid and liquid parts on each other. An ALE formulation through a Moving Mesh node is used to observe geometric changes of the fluid of the flapping motion. This study depends on time to simulate the flapping of the wing along the air flow.

\subsection {Materials and Methods}

The importance of flapping wing micro aircraft has increased in recent years. With the increase in importance, studies have intensified to design better micro aircraft. Engineers and scientists inspired by nature in their works. Moths, cicadas, hummingbirds etc. from flying creatures inspired by nature have been the subject of research for micro aircraft. Analyses are carried out to observe the forces these flying creatures produce during flight. It cannot be adequately represented by 2D simulations for flapping wing analysis. 

\begin{figure}[ht]
    \centering
    \includegraphics[width=0.45\textwidth]{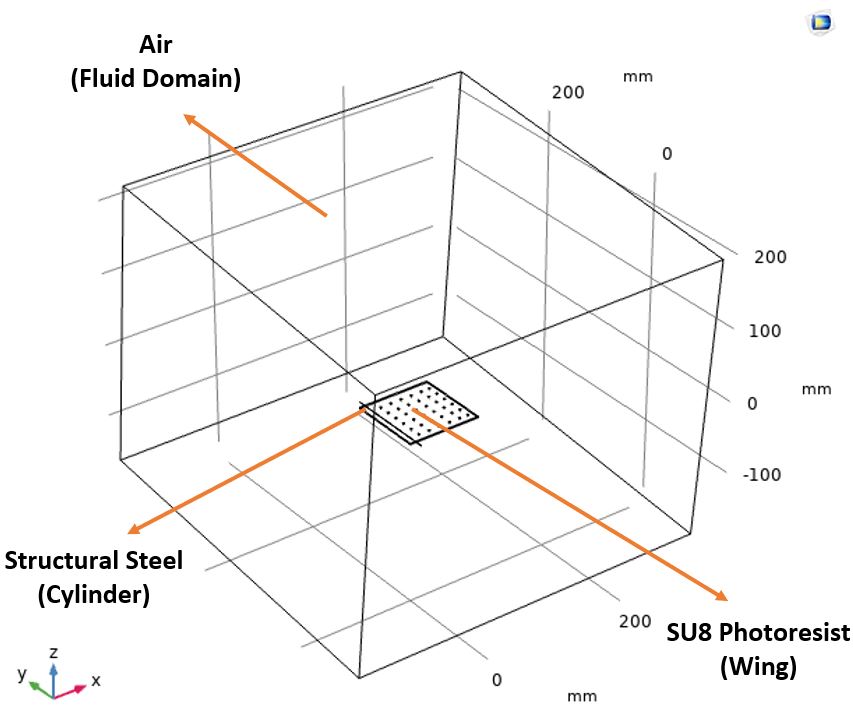}
    \caption{Materials of the Parts}
    \label{fig:p2}
\end{figure}

In this study, the geometry is formed by assembling a wing and a cylinder. In order to represent the axis of rotation of the wing, the cylinder was added to the wing body. Since the body of the wing is flexible, SU8 - photoresist material which is flexible has been assigned to the wing. Since the cylindrical body around which the wing rotates is not included in the force calculations, structural steel has been assigned as a rigid material. In order to perform the wind tunnel test in the simulation, the air was assigned as a rectangular prism around the geometry of this wing and the cylinder assembling.In this analysis model assemblies materials are combination of flexible and rigid components in this reason the Multibody Dynamics interface was used in the Comsol Multiphysics . In order to model the flapping wing mechanism, the joints, attachments etc. properties of this interface were used. These parts of geometry as shown in Figure \ref{fig:p2}. 

\subsection {Moving Mesh}

\begin{figure}[ht]
    \centering
    \includegraphics[width=0.42\textwidth]{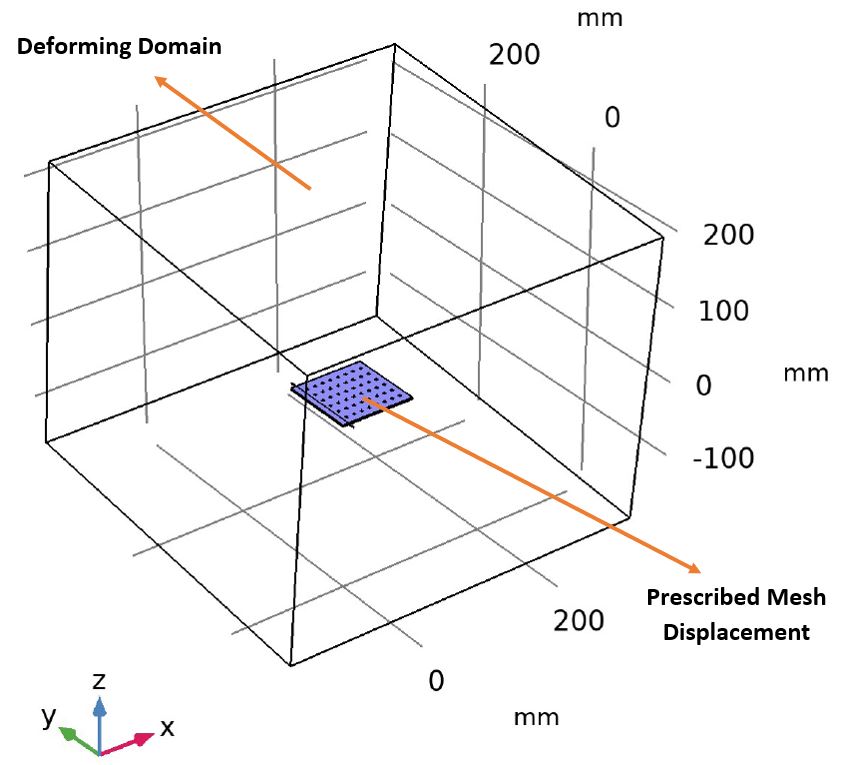}
    \caption{Transferring Solid Motion to Moving Mesh Boundaries}
    \label{fig:p3}
\end{figure}

Moving mesh uses to study the geometry shape is changed due to the movement of the solid boundaries and the deformation of the solid areas. It can be used for time-dependent and stationary situations. The flapping wing movement in the air domain modeled using an ALE formulation. A deforming domain condition is assigned to the fluid area at the moving boundaries. In our study, a Prescribed Mesh Displacement mesh boundary condition is used to realize the motion of the mesh. Thus, the motion of the wing has been transferred to the moving mesh at all solid-fluid boundaries on the wing. As shown in Figure \ref{fig:p3}, this boundary condition is assigned to the solid boundary in the area where the wing and fluid intersect.

\subsection {Fluid Flow}
Fluid in a wind tunnel is defined by in-compressible Navier-Stokes equations for the velocity field and pressure in the deformed spatial coordinate system. In other words, this formulation allows to describe the motion of fluids such as liquids and gases. When using a compressible fluid, the following equations are used:
\begin{equation}
\begin{array}{l}
\scalebox{1}{%
$ 
\rho \frac{\partial \mathbf{u}_{\text {fluid }}}{\partial t}+\rho\left(\mathbf{u}_{\text {fluid }} \cdot \nabla\right) \mathbf{u}_{\text {fluid }}=\nabla \cdot[-p \mathbf{l}+\mathbf{K}]+\mathbf{F} $}
\end{array}
\end{equation}

\begin{equation}
\begin{array}{l}
\scalebox{1}{%
$ 
\rho \nabla \cdot\left(\mathbf{u}_{\text {fluid }}\right)=0 $}
\end{array}
\end{equation}

\begin{equation}
\begin{array}{l}
\scalebox{1}{%
$ 
\mathbf{K}=\mu\left(\nabla \mathbf{u}_{\text {fluid }}+\left(\nabla \mathbf{u}_{\text {fluid }}\right)^{\top}\right) $}
\end{array}
\end{equation}
where u is the fluid velocity, p is the fluid pressure, {$\rho$} is the fluid density, {$\mu$} is the fluid dynamic viscosity and T is temperature.

The Navier-Stokes equations symbolize the conservation of momentum by the other hand the continuity equation symbolize the conservation of mass.

\subsection {Numerical Method}
The following equations are used to calculate the lift and drag forces produced by the flapping wing.

Lift Force is:
\begin{equation}
\begin{array}{l}
\scalebox{1}{%
$ 
F_{L}=C_{L} \frac{1}{2} \rho A U^{2} $}
\end{array}
\end{equation}

Drag Force is:
\begin{equation}
\begin{array}{l}
\scalebox{1}{%
$ 
F_{D}=C_{D} \frac{1}{2} \rho A U^{2} $}
\end{array}
\end{equation}

Analyses were performed using these equations at 25 Hz flapping frequency and the flapping angle of the wing was given as 30 degrees.

\begin{figure}[ht]
    \centering
        \includegraphics[width=0.5\textwidth]{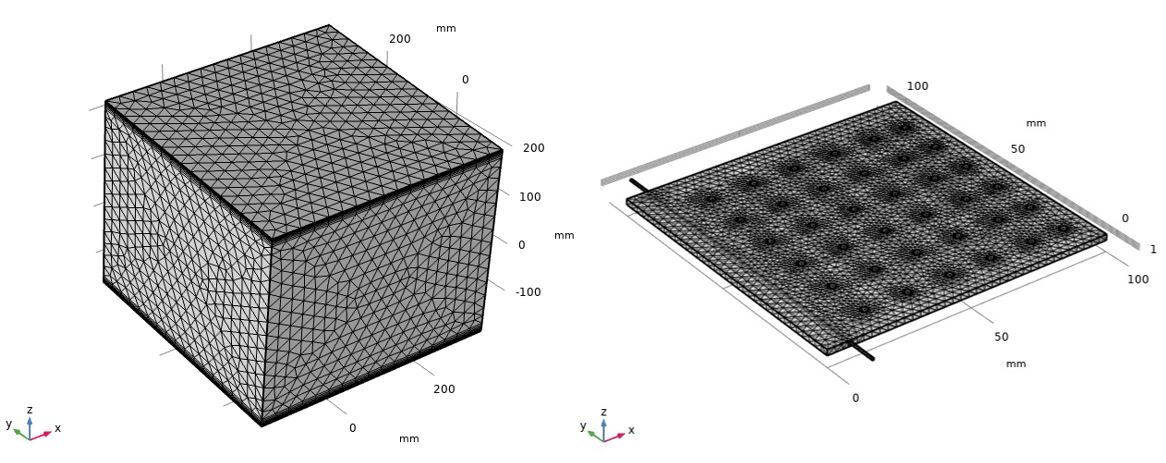}
    \caption{Mesh on the Air Domain, Wing and Cylinder}
    \label{fig:p4}
\end{figure}

Element is the name given to each of the components of the mathematical model that are divided in such a way that they do not overlap. The function to be calculated is applied to each of the divided parts. At the end, these parts are brought together to obtain an approximate result. This is done by different meshing methods such as tetrahedra (tets), hexahedra (bricks), triangular prisms (prisms) and pyramids. The meshes of the air domain, wing and cylinder are shown in Figure \ref{fig:p4}. These mesh methods change with the angle of inclination and numbers of the pores. Air was supplied to the air domain inlet region at a speed of 5 m/s.

\section{Simulation Results}

In this work, analyses of the wing were made at different pore and inclination angles using Comsol Multiphysics. As a result of the analysis, the average lift values are calculated for the whole wing.

In the analyses of the at different degrees inclined porous of wing, the number of pores was increased and the effect of this situation on the lift and drag forces was observed. Comparison was made by taking the average of the lift values of the analyses. As a result of the comparisons made, as the number of pores on the wing increases, the lift value moves away from zero. Lift forces according to pore numbers and angle of inclinations with nonporous result are shown in Table \ref{tab:lift_result}.

\begin{table}[ht]
\renewcommand{\arraystretch}{1.6}
\centering
\caption{Lift Force Results}
 \begin{tabular}{|| c c c c||}
 \hline \hline
      &  60 Degree & 70 Degree & 80 Degree \\
 \hline
  36 pores  &  -0,001952 & -0,000984 & -0,000504 \\
\hline
  48 pores  & -0,00217 & -0,001496 & -0,0011436   \\
\hline
  60 pores  & -0,00116 & -0,001826 & -0,001272 \\
\hline\hline\hline
 Nonporous   & -0,0015354 &  &  \\
\hline
\end{tabular}
\label{tab:lift_result}
\end{table}

Analyses were started to be made by changing the pore numbers keeping the angles of inclination degrees constant. In the analysis of the 60 degree inclined porous wing, the pore number was increased by keeping the angle of inclination variable constant, and the effect of this situation on the lift and drag forces was observed. Comparison was made by taking the average of the lift values of the analyses. The average lift values' result is negative. As a result of comparisons, as the number of pores on the wing increases, the lift value moves away from zero. After 60 degree analyses finished, 70 degree wing analyses were performed by repeating the same procedures. In the results of these analyses the lift values were calculated, it was observed that they gave results compatible with the first analyses. After these analyses, finally 80 degree analyses were performed. It was observed that the results of the last analyses obtained were compatible with the results of other analyses.

While the different degree angle of inclination of the pores is constant, the lift graphics calculated by changing the number of pores are similar. Also when the number of pores is constant, the lift graphics calculated by changing the different degree angle of inclination of the pores are similar with each others. Although the up and down flapping angle, which is 30 degree, of the wing is equal, the resulting lift waves are not symmetrical. The values occurring in the down-stroke of the wing are numerical value greater than the values occurring in the up-stroke. 

\section{Conclusion}

In this study, when the wing surface has a porous structure, the effect of this on the lift and drag forces which generated during flight was calculated. The research was performed numerically, by simulating laminar flow of air flows acting on porous wing. The graphs for the calculation of the lift force are as oscillation. The negative peaks of these plots are bigger than the positive peaks. The result of this, the average of lift is negative. Due to the pores in the wing, the upper and lower wing surface areas are different from each other. Also, the minus average lift value observed as a result of the nonporous wing analysis is explained by the small angle of attack. Since the upper plane area of the wing is smaller than the downstream area, the total lift produced during flapping is less than the total lift produced during landing. In addition, another factor that makes the lift negative is the small angle of attack which is 5 degree at this analyses. Thus, the net lift is negative. From these results, it has been observed that the pores on the wing affect the forces.

\bibliographystyle{ieeetr}
\bibliography{kaynakca.bib}

\begin{thebibliography}{10}

\bibitem{mcmichael1997micro}
J.~M. McMichael, ``Micro air vehicles-toward a new dimension in flight,'' {\em
  http://www. arpa. gov/tto/MAV/mav\_auvsi. html}, 1997.

\bibitem{Yang2018THETF}
L.-J. Yang, A.-L. Feng, H.~Lee, B.~Esakki, and W.~He, ``The three-dimensional
  flow simulation of a flapping wing,'' {\em Journal of Marine Science and
  Technology}, vol.~26, pp.~297--308, 2018.

\bibitem{moses2017artificial}
K.~Moses, S.~Michaels, M.~Willis, and R.~Quinn, ``Artificial manduca sexta
  forewings for flapping-wing micro aerial vehicles: how wing structure affects
  performance,'' {\em Bioinspiration \& biomimetics}, vol.~12, no.~5,
  p.~055003, 2017.

\bibitem{deng2006flapping}
X.~Deng, L.~Schenato, W.~C. Wu, and S.~S. Sastry, ``Flapping flight for
  biomimetic robotic insects: Part i-system modeling,'' {\em IEEE Transactions
  on Robotics}, vol.~22, no.~4, pp.~776--788, 2006.

\bibitem{azuma1985flight}
A.~Azuma, S.~Azuma, I.~Watanabe, and T.~Furuta, ``Flight mechanics of a
  dragonfly,'' {\em Journal of experimental biology}, vol.~116, no.~1,
  pp.~79--107, 1985.

\bibitem{sunada1993performance}
S.~Sunada, K.~Kawachi, I.~Watanabe, and A.~Azuma, ``Performance of a butterfly
  in take-off flight,'' {\em Journal of Experimental Biology}, vol.~183, no.~1,
  pp.~249--277, 1993.

\bibitem{wood2008first}
R.~J. Wood, ``The first takeoff of a biologically inspired at-scale robotic
  insect,'' {\em IEEE transactions on robotics}, vol.~24, no.~2, pp.~341--347,
  2008.

\bibitem{singh2008insect}
B.~Singh and I.~Chopra, ``Insect-based hover-capable flapping wings for micro
  air vehicles: experiments and analysis,'' {\em AIAA journal}, vol.~46, no.~9,
  pp.~2115--2135, 2008.

\bibitem{phan2017design}
H.~V. Phan, T.~Kang, and H.~C. Park, ``Design and stable flight of a 21 g
  insect-like tailless flapping wing micro air vehicle with angular rates
  feedback control,'' {\em Bioinspiration \& biomimetics}, vol.~12, no.~3,
  p.~036006, 2017.

\bibitem{nguyen2015performance}
Q.~Nguyen, W.~Chan, and M.~Debiasi, ``Performance tests of a hovering flapping
  wing micro air vehicle with double wing clap-and-fling mechanism,'' in {\em
  International Micro Air Vehicles Conference and Flight Competition},
  pp.~1--8, 2015.

\bibitem{bruggeman2010improving}
B.~Bruggeman, ``Improving flight performance of delfly ii in hover by improving
  wing design and driving mechanism,'' {\em Delft University of Technology M.
  Sc. thesis}, 2010.

\bibitem{rose2015modeling}
C.~J. Rose, {\em Modeling and Control of an Ornithopter for Non-Equilibrium
  Maneuvers}.
\newblock PhD thesis, UC Berkeley, 2015.

\bibitem{pornsin2000microbat}
T.~Pornsin-Sirirak, Y.~Tai, C.~Ho, and M.~Keennon, ``Microbat: A palm-sized
  electrically powered ornithopter. 2001 nasa,'' in {\em JPL Workshop on
  Biomorphic Robotics, Pasadena, CA}, 2000.

\bibitem{yang2012micro}
L.-J. Yang {\em et~al.}, ``The micro-air-vehicle golden snitch and its
  figure-of-8 flapping,'' {\em Journal of Applied Science and Engineering},
  vol.~15, no.~3, pp.~197--212, 2012.

\bibitem{ramezani2016bat}
A.~Ramezani, X.~Shi, S.-J. Chung, and S.~Hutchinson, ``Bat bot (b2), a
  biologically inspired flying machine,'' in {\em 2016 IEEE International
  Conference on Robotics and Automation (ICRA)}, pp.~3219--3226, IEEE, 2016.

\bibitem{roshanbin2017colibri}
A.~Roshanbin, H.~Altartouri, M.~Kar{\'a}sek, and A.~Preumont, ``Colibri: A
  hovering flapping twin-wing robot,'' {\em International Journal of Micro Air
  Vehicles}, vol.~9, no.~4, pp.~270--282, 2017.

\bibitem{corporate2012smartbird}
F.~Corporate, ``Smartbird--bird flight deciphered,'' {\em Retrieved July},
  2012.

\bibitem{willmott1997mechanics}
A.~P. Willmott and C.~P. Ellington, ``The mechanics of flight in the hawkmoth
  manduca sexta. i. kinematics of hovering and forward flight.,'' {\em Journal
  of experimental Biology}, vol.~200, no.~21, pp.~2705--2722, 1997.

\bibitem{wu2011structural}
P.~Wu, B.~Stanford, E.~S{\"a}llstr{\"o}m, L.~Ukeiley, and P.~Ifju, ``Structural
  dynamics and aerodynamics measurements of biologically inspired flexible
  flapping wings,'' {\em Bioinspiration \& biomimetics}, vol.~6, no.~1,
  p.~016009, 2011.

\bibitem{pelletier2000low}
A.~Pelletier and T.~J. Mueller, ``Low reynolds number aerodynamics of
  low-aspect-ratio, thin/flat/cambered-plate wings,'' {\em Journal of
  aircraft}, vol.~37, no.~5, pp.~825--832, 2000.

\bibitem{mazaheri2010experimental}
K.~Mazaheri and A.~Ebrahimi, ``Experimental investigation of the effect of
  chordwise flexibility on the aerodynamics of flapping wings in hovering
  flight,'' {\em Journal of Fluids and Structures}, vol.~26, no.~4,
  pp.~544--558, 2010.

\bibitem{kumar2016analysis}
A.~Kumar, C.~Kaur, and S.~S. Padhee, ``Analysis and optimization of dragonfly
  wing,'' in {\em Proceedings of the 2016 COMSOL Conference (User
  Presentations) in Bangalore, India}, 2016.

\end{thebibliography}

\end{document}